\documentclass[12pt]{article}

\usepackage{amsmath}
\usepackage{amssymb}
\usepackage{multirow}
\usepackage{color}
\renewcommand{\epsilon}{\varepsilon}

\usepackage{amsmath}
\usepackage{amssymb}
\usepackage[ansinew]{inputenc}
\usepackage{graphicx}
\usepackage{epsfig}

\usepackage[authoryear]{natbib}
\newtheorem{theorem}{Theorem}[section]

\newtheorem{lemma}{Lemma}[section]
\newtheorem{example}{Example}[section]
\newtheorem{remark}{Remark}[section]

\def\IM{{\bf I\kern-.25em M}}

\def\3{\ss}
\def\er{\mathbb{R}}

\newcommand{\bea}{\begin{eqnarray*}}
\newcommand{\eea}{\end{eqnarray*}}
\newcommand{\be}{\begin{eqnarray}}
\newcommand{\ee}{\end{eqnarray}}

\newcommand{\ba}{\begin{array}}
\newcommand{\ea}{\end{array}}
\def\3{\ss}
\def\er{\mathbb{R}}

\newcommand{\ve}{\varepsilon}

\newcommand{\btheta}{\theta}

\title{A new  approach to optimal designs for  correlated observations}

\begin{document}

\author{Holger Dette, Maria Konstantinou    \\
Ruhr-Universit\"at Bochum \\
Fakult\"at f\"ur Mathematik \\
44780 Bochum \\
Germany
\and
Anatoly Zhigljavsky \\
School of Mathematics\\
 Cardiff  University\\
  Cardiff, CF24 4AG\\
  UK
}
\date{}
\maketitle

\begin{abstract}

This paper presents a new and efficient method for the construction of  optimal designs for regression models with dependent error processes.
In contrast to most of the work in this field, which starts with a model for a finite number of observations and considers the asymptotic properties
of estimators and designs as the sample size converges to infinity, our approach
is based on a continuous time model.
We use results from    stochastic analysis to identify the
best linear unbiased estimator (BLUE) in this model. Based on the BLUE, we construct an efficient linear estimator and corresponding optimal designs in the model for
finite sample size by minimizing the mean squared error between the optimal solution in the continuous time model and its
discrete approximation with respect to the weights (of the linear estimator) and the optimal design points, in particular in the multi-parameter case. \\
In contrast to previous work on the subject the resulting estimators and corresponding optimal designs are very
 efficient and easy to implement. This means that they are practically not distinguishable from the weighted least squares estimator and the corresponding
optimal designs, which have to be found numerically by non-convex discrete optimization.
The advantages of the new approach  are illustrated in    several  numerical examples.
\end{abstract}



\section{Introduction}
\def\theequation{1.\arabic{equation}}
\setcounter{equation}{0}

 The construction of optimal designs for dependent observations is a very
 challenging problem in statistics, because - in contrast to the independent case -
 the dependency  yields  non-convex optimization  problems.
 As a consequence, classical tools of convex optimization theory
 as described, for example, in \cite{pukelsheim2006} are not applicable.
 Most of the discussion is restricted to very simple models  and we refer
to \cite{detkunpep2008,KisStehlik2008,harstu2010}  for some
exact optimal designs for linear regression models.
Several authors have proposed to determine
optimal designs using asymptotic
arguments [see, for example,  \cite{sackylv1966,sackylv1968}, \cite{bickherz1979}, \cite{N1985a}, \cite{zhidetpep2010}],
but  the  resulting approximate  optimal design problems are still  non-convex and extremely difficult to solve.
As  a consequence,  approximate optimal designs have mainly  been determined  analytically for the location  model
(in this case the corresponding optimization problems are in fact convex) and  for a few one-parameter linear models [see
\cite{bolnat1982}, \cite{N1985a}, Ch.\ 4, \cite{naether1985b},
 \cite{pazmue2001}   and  \cite{muepaz2003} among others].  \\
Recently, substantial progress has been made in the construction of optimal designs for regression models with a dependent error process.
\cite{DetPZ2012} determined (asymptotic) optimal designs for least squares estimation,  under the additional assumption
that the regression functions are eigenfunctions of an integral operator associated with the covariance kernel of the error process.
Although  this approach is able to deal with the multi-parameter case,  the class of models for which approximate optimal designs can be determined explicitly
 is still rather small, because it refers to   specific kernels with corresponding eigenfunctions.
 For this reason \cite{detpepzhi2015}
proposed a different strategy to obtain optimal designs and efficient estimators. Instead of constructing an optimal design for a particular estimator (such as least squares or weighted least squares), these authors proposed to consider the problem of optimizing the estimator and the design of experiment simultaneously. They constructed a class of estimators and corresponding optimal designs with a variance converging (as the sample size increases) to the optimal variance in the continuous time model. In other words, asymptotically these estimators achieve the same precision as the best linear unbiased estimator computed from the whole trajectory of the process. While this approach yields a satisfactory solution for one-dimensional parametric models using signed least squares estimators, it is not transparent and in many cases not efficient
 in the multi-parameter model. In particular, it is based on matrix-weighted linear estimators and corresponding designs  which are  difficult to implement  in practice and do not yield the same high efficiencies as in the one-dimensional case. \\
 In this paper  we present an alternative approach  for the construction of estimators and corresponding
optimal designs for regression models with dependent error processes, which
has  important advantages compared  to the currently used methodology.
First - in contrast to all other  methods  -  the estimators with corresponding optimal designs   proposed  here are
 very  easy to implement. Secondly, it is demonstrated that   the new estimator and design yield    a method which is practically not distinguishable from the best
 linear estimator (BLUE) with corresponding optimal design. Third, in many cases the new estimator and a uniform design are already very efficient.  \\
Compared to most of the work in this field, which begins with a model for a finite number of observations and
 considers the asymptotic properties
of estimators as the sample size converges to infinity, an essential difference of our  approach is that it is directly based  on   the continuous time model.
In Section \ref{sec2} we derive the best linear unbiased estimate in this model
  using  results about the absolute continuity of measures on the space $C([a,b])$. This yields a representation of the best linear estimator as a stochastic integral and provides
an efficient tool for constructing estimators with corresponding optimal designs for finite samples
which are practically not distinguishable from the optimal (weighted least squares) estimator and corresponding optimal design. We emphasize again that the latter design has to be determined by discrete non-convex optimization. To be more precise, in Section \ref{sec3} we propose a weighted mean, say $\sum^n_{i=1}\mu_i Y_{t_i}$ (here $Y_{t_i}$ denotes the response at the point $t_i$ and $n$ is the sample size),
where the weights $\mu_{1},\dots,\mu_n$ (which are vectors in case of models with more than one parameter) and  design points $t_1,\dots,t_n$ are determined
 by minimizing the mean squared error between the optimal solution in the continuous time model (represented by a stochastic integral with respect to the underlying process) and its
discrete approximation with respect to the weights (of the linear estimator) and the optimal design points.
In  Section \ref{sec4} we discuss several examples and demonstrate the superiority of the new approach
to the method which was recently proposed in  \cite{detpepzhi2015}, in particular for multi-parameter models.
Some more details  on best linear unbiased estimation in the continuous time model
are given  in Section \ref{secextra}, where we discuss  degenerate cases, which appear - for example - by a constant term in the regression
function.  For a more transparent presentation of the ideas some technical details are additionally deferred to the Appendix. \\
We finally note that this paper is a first approach which uses results from  stochastic analysis in the context of optimal
design theory. The combination of these two fields yields a practically implementable and satisfactory solution of optimal design problems
for a broad class of regression models with dependent observations.

\section{Optimal estimation in continuous time models} \label{sec2}
\def\theequation{2.\arabic{equation}}
\setcounter{equation}{0}

Consider a   linear regression model of the form
\be
\label{eq:model}
 Y_{t_i}=  Y(t_i)=\theta^T f(t_i)+\varepsilon_{t_i} ,~~\quad  i=1,\dots,n \, ,
     \ee
     where $\{  \varepsilon_{t} \mid t \in [a,b] \}$ is a Gaussian process,
     $     \mathbb{E}[\varepsilon_{t_i} ]=0, $ $K(t_i,t_j) = \mathbb{E}[\varepsilon_{t_i}\varepsilon_{t_j}]$ denotes the covariance between observations at the points $t_i$ and $t_j$  ($ i,j=1,\dots,n $),  $\theta=(\theta_1, \ldots, \theta_m)^T$ is a vector of unknown parameters, $f(t)=(f_1(t), \ldots, f_m(t))^T$ is a vector of continuously differentiable linearly independent functions, and the explanatory variables $t_1,\dots,t_n$  vary in a compact interval, say $[a,b]$.  If   $\mathbf{Y}=(Y_{t_1}, \ldots, Y_{t_n})^T$  denotes the
     vector of observations the  weighted  least squares estimator   of $\theta$ is defined by
  \begin{equation*}
  \widehat \btheta_{WLSE}  = (\mathbf{X}^T\mathbf{\Sigma}^{-1}\mathbf{X})^{-1}\mathbf{X}^T\mathbf{\Sigma}^{-1} \mathbf{Y},
\end{equation*}
where $\mathbf{X}=(f_p(t_j))^{p=1,\ldots,m}_{j=1,\ldots,n}$ is the $n\!\times\! m$ design matrix and
 $\mathbf{\Sigma}=(K(t_i,t_j))_{i,j=1,\ldots,n}$ is the $n\!\times\! n$ matrix of variances/covariances. It is well known that $\hat \theta_{WLSE}$ is the BLUE in model \eqref{eq:model}.
 The corresponding minimal variance is given by
\be
 \mathrm{Var} (\widehat\btheta_{WLSE}) &=~
 (\mathbf{X}^T \mathbf{\Sigma}^{-1}\,\mathbf{X})^{-1}  \label{eq:var-wls},
\ee
and an optimal design for the estimation of the parameter $\theta$ in  model \eqref{eq:model}  minimizes an appropriate real-valued functional of this matrix.
As pointed out before, the direct minimization   of this type of criterion is an extremely challenging non-convex discrete optimization problem and explicit solutions are not available in  nearly all cases of practical interest. For this reason many authors propose to consider asymptotic optimal designs as the sample size $n$ converges to infinity [see   \cite{sackylv1966,sackylv1968}, \cite{bickherz1979}, \cite{N1985a}, \cite{zhidetpep2010}]. \\
In the following discussion we consider - parallel to model \eqref{eq:model} -  its  {\it continuous time} version, that is
\be \label{mod1cont}
Y_t = \theta^Tf(t)+ \varepsilon_{t} ~,~~ t \in [a,b],
\ee
where the full trajectory of the process $\{Y_t |  \ t\in[a,b]\}$ can be observed and $\{\varepsilon_{t} |  \ t\in[a,b]\}$  is a centered Gaussian process
 with continuous  covariance kernel $K$, i.e. $K(t,t') = \mathbb{E}[\varepsilon_{t} \varepsilon_{t'} ]$.  We will focus on
  {\it triangular kernels}, which are of the form
\be
\label{eq:cov_tr0}
K(t,t')=u(t)v(t')\;\;\; {\rm for }\;\; t \leq t',
\ee
$(K(t,t')=K(t',t)$ for $t > t'$),
where $u(\cdot)$ and $v(\cdot)$ are  some  functions  defined on the interval
$[a,b]$. An alternative representation of $K$  is
given by
$$K(t,t')=v(t) v(t') \min\{q(t),q(t')\}; \quad (t, t' \in [a,b]),$$
where  $q (t)= u(t)/v(t)$.  We assume that the process
$\{ \ve_t  | ~ [a,b]\}$ is non-degenerate on the open interval $(a,b)$, which implies that
the function $q$ is positive  on the interval $(a,b)$ and
strictly increasing and continuous on $[a,b]$,  see \cite{mehr1965certain} for more details.
Consequently,   the functions  $u$ and $v$ must   have the same sign and
can be assumed to be  positive on the interval $(a,b)$ without loss of generality.
Note that the majority of covariance kernels considered in the literature belong to this class, see,
 for example, \cite{N1985a,zhidetpep2010} or \cite{Harman}.
 The simple triangular kernel\begin{equation*}
K(t,t')= t \wedge t',
\end{equation*}
 is obtained for the choice $u(t)=t$ and $v(t)=1$ and corresponds to the  Brownian motion. As pointed out
  in  \cite{detpepzhi2015}, the solutions of the optimal design problems with  respect to different
  triangular kernels are closely related. In particular, if a best linear unbiased estimator (BLUE)  for a particular triangular kernel
  has to be found for the continuous time model, it can be obtained by simple nonlinear transformation from the BLUE in a different continuous time model (on a possibly different interval) with a Brownian motion as error process  (see Remark \ref{rem1}(b)  below for more details). For this reason we will concentrate on the covariance
   kernel of the Brownian motion throughout this section.
Our first  result provides the optimal estimator in the continuous time model \eqref{mod1cont}, where the error process is given
by a Brownian motion on the interval $[a,b]$, where $a>0$ (the case $a=0$ will be discussed in Section \ref{secextra}).
We begin with a lemma which is crucial for the definition
of the estimator. The proof can be found in the Appendix.

\begin{lemma} \label{lemma0} Consider the continuous time linear regression model \eqref{mod1cont} on the interval $[a,b]$, $a >0$,
with a continuously differentiable vector of regression
functions $f$ and a Brownian motion as error process.  Then
the $m \times m$ matrix
\begin{equation}\label{cmatrix}
C = \int_a^b \dot{f}(t) \dot{f}^T(t) \,dt + \frac{f(a) f^T(a)}{a}
\end{equation}
is non-singular.
\end{lemma}

\begin{theorem}
\label{thm1}
Consider the continuous time linear regression model \eqref{mod1cont} on the interval $[a,b]$, $a >0$,
with a continuously differentiable vector of regression
functions $f$ and a Brownian motion as error process.  The best linear unbiased estimate is given by
\begin{equation}\label{asblue}
\hat{\theta}_{\rm BLUE}   = C^{-1} \Big( \int_a^b \dot{f}(t) \,dY_t + \frac{ f(a)}{a} Y_a \Big) .
\end{equation}
Moreover, the minimum variance is given by
\begin{equation}\label{cmatrix1}
C^{-1}  = \Big(  \int_a^b \dot{f}(t) \dot{f}^T(t) \,dt + \frac{f(a) f^T(a)}{a} \Big)^{-1} .
\end{equation}
\end{theorem}

{\bf Proof of Theorem \ref{thm1}.}   Note that  the continuous time model \eqref{mod1cont}
can be written as a Gaussian white noise model
\begin{equation*}
Y_t  = \int_0^t s_1(u) \,du +  \int_0^t \, d \varepsilon_u, \quad t \in [0,b] ,
\end{equation*}
where the function $s_1$ is defined as
\begin{equation*}
s_1(u) = I_{[a,b]}(u) \theta^T \dot{f}(u) + I_{[0,a]}(u) \frac{\theta^T f(a)}{a} .
\end{equation*}
Let $\mathbb{P}_\theta$  and $\mathbb{P}_0$  denote the measure on $C([0,b])$ associated with the process
$Y = \{ Y_t | \ t \in [0,b] \}$ and $\{ \varepsilon_{t} | \ t \in [0,b] \}$, respectively. From Theorem 1 in Appendix II of
\cite{MR620321} it follows that $\mathbb{P}_1$ is absolute continuous with respect to $\mathbb{P}_2$ with
Radon-Nikodym derivative given by
\begin{align*}
\frac{d \mathbb{P}_\theta}{ d \mathbb{P}_0} (Y) &= \exp \left\{  \int_0^b s_1(t) \,dY_t - \frac{1}{2} \int_0^b s_1^2(t) \,dt \right\} \\ &=
\exp \left\{ \left(\int_a^b \theta^T \dot{f}(t) \,dY_t + \frac{\theta^T f(a)}{a} Y_a \right) - \frac{1}{2} \left( \int_a^b (\theta^T \dot{f}(t))^2 \,dt + \frac{(\theta^T f(a))^2}{a} \right)  \right\} .
\end{align*}
The maximum likelihood estimator can be determined by  solving  the  equation
\begin{equation*}
\frac{\partial}{\partial \theta} \log \frac{d \mathbb{P}_\theta}{ d \mathbb{P}_0} (Y) =
 \int_a^b \dot{f}(t) \,dY_t + \frac{ f(a)}{a} Y_a  -  \Big( \int_a^b \dot{f}(t) \dot{f}^T(t) \,dt + \frac{f(a) f^T(a)}{a} \Big) \theta = 0.
\end{equation*}
The solution coincides with  the linear estimate \eqref{asblue}, and a straightforward calculation, using Ito's formula
and the fact that the random variables $\int^b_a \dot{f} (t) d \varepsilon_t $ and $ \varepsilon_a$ are independent, gives
\begin{eqnarray*}
 \mbox{Var}_\theta(\hat{\theta}_{\rm BLUE} ) &=& C^{-1} \mathbb{E}_{\theta} \Bigl [ \Big ( \int^b_a \dot{f} (t) d \varepsilon_t + \frac {f(a)}{a}
 \varepsilon_a \Big) \Big( \int^b_a \dot{f} (t) d \varepsilon_t + \frac {f(a)}{a} \varepsilon_a \Big)^T \Bigr] C^{-1} \\
&=& C^{-1} \Big ( \int^b_a \dot{f} (t)  \dot{f}^T (t) dt + \frac {f(a) f^T(a)}{a} \Big) C^{-1} = C^{-1},
  \end{eqnarray*}
where the matrix $C$ is defined in \eqref{cmatrix}. It has been shown in \cite{detpepzhi2015} that this matrix is the
variance/covariance matrix of the BLUE in the continuous time model, which  proves Theorem \ref{thm1}.
\hfill $\Box$

\begin{remark} \label{rem1} {\rm ~ \\
\vspace{-.75cm}
\begin{itemize}
\item[(a)]
\cite{detpepzhi2015}  determined the best linear estimator for the continuous time linear regression model \eqref{mod1cont} with a twice continuously differentiable
vector of regression functions and Brownian motion as
\begin{equation}
\label{BLUE_DPZ}
C^{-1} \Big  \{ \dot{f}  (b) Y_b + \Big ( \frac {f(a)}{a} - \dot{f} (a) \Big ) Y_a - \int^b_a \ddot{f} (t) Y_t dt \Big \}.
\end{equation}
Using integration by parts gives
\begin{equation*}
\int_a^b \dot{f}(t) \,dY_t = \dot{f}(b) Y_b - \dot{f}(a) Y_a - \int_a^b \ddot{f}(t) Y_t \,dt ,
\end{equation*}
 and it is easily seen that the expression \eqref{BLUE_DPZ} coincides
with \eqref{asblue}.  This means  that a BLUE in the continuous time model \eqref{mod1cont}  is even available under
the weaker assumption of a once   continuously differentiable function $f$.
\item[(b)] The best linear estimator in the continuous time model \eqref{mod1cont} with a general triangular kernel of the form
\eqref{eq:cov_tr0} can easily be obtained  from Appendix $B$ in \cite{detpepzhi2015}. To be precise, consider a triangular kernel of the form \eqref{eq:cov_tr0},
define
\begin{align*}
 q(t) = \frac{u(t)}{v(t)}, ~
 \alpha(t) = v(t) ,
\end{align*}
and consider the stochastic process
\begin{equation*}
\varepsilon_t = \alpha(t) \tilde{\varepsilon}_{q(t)} ,
\end{equation*}
where $\{ \tilde{\varepsilon}_{\tilde{t}} | \ \tilde{t} \in [\tilde{a},\tilde{b}] \}$  is a Brownian motion
on the interval $[\tilde{a},\tilde{b}]$ and  $\tilde{a} = q (a)$, $\tilde{b} =  q (b)$.  It follows
from  \cite{doob1949heuristic}
that  $\{ \varepsilon_t  | \ t \in [a,b] \}$  is a centered Gaussian process on the interval $[a,b]$ with covariance kernel
\eqref{eq:cov_tr0}. Moreover, if we consider the continuous time model
\begin{equation}\label{transformed-model}
\tilde Y_{\tilde t} = \theta^T \tilde{f}(\tilde{t}) + \tilde{\varepsilon}_{\tilde{t}}, \quad \tilde{t} \in [\tilde{a}, \tilde{b}],
\end{equation}
and use the transformations
\be
\label{trans}
\tilde{f}(\tilde{t}) =  \frac{f(q^{-1}(\tilde{t}))}{v(q^{-1}(\tilde{t}))}~,~
 \tilde{\varepsilon}_{\tilde{t}}  = \frac{\varepsilon_{q^{-1}(\tilde{t})}}{v(q^{-1}(\tilde{t}))} ~,~\tilde Y_{\tilde t}  = \frac{Y_t}{v(t)},
\ee
then it follows from \cite{detpepzhi2015}  that the BLUE  for
the   continuous time model \eqref{mod1cont} (with a general triangular covariance kernel) can be obtained from the
BLUE in  model \eqref{transformed-model} by the transformation
$\tilde{t} = q (t)$. Therefore an application of Theorem \ref{thm1} gives for the best linear estimator in the continuous time model \eqref{mod1cont}
with triangular covariance kernel of the form \eqref{eq:cov_tr0} the representation
\begin{equation*}
\hat{\theta}_{\rm BLUE} = C^{-1} \Big [ \int_a^b \frac{\dot{f}(t)  v(t) - {\dot{v}(t)}  f(t)}{\dot{u}(t)v(t)-u(t)\dot{v}(t)} \,d \left(\frac{Y_t}{v(t)} \right) + \frac{f(a)}{u(a)v(a)} Y_a \Big] ~,
\end{equation*}
where the matrix $C$ is given by
\begin{equation*}
C = \int_a^b \frac{ [ \dot{f}(t) v(t) -   \dot{v}(t) \dot{f}(t)]  [  \dot{f}(t) v(t) -   \dot{v}(t) \dot{f}(t) ] ^T }
{v^2(t) [ \dot{u}(t)v(t)-u(t)\dot{v}(t)] } \,dt + \frac{f(a)f^T(a)}{u(a)v(a)}.
\end{equation*}

\item[(c)]    Using integration by parts it follows (provided that the functions $f$, $u$, and $v$
are twice continuously differentiable)  that  the BLUE in
the continuous time model \eqref{mod1cont}  can be represented as
\begin{equation*} 
\hat \theta_{\rm BLUE} = \int_a^b Y_t \, \mu^*(dt) ,
\end{equation*}
where $\mu^*$ is a vector of signed measures defined by $ \mu^*(dt)  =  P_a \delta_a  + p(t) dt + P_b \delta_b $, $\delta_t$ denotes the  Dirac
measure at the point $t \in [0,1]$ and the  ``masses'' $P_a$, $P_b$ and the density $p$ are given by
\begin{eqnarray*}
P_a
&=&
C^{-1}
 \frac{1}{u(a)} \frac{  f(a)  \dot{u}(a)  - \dot f (a) u(a)   }{  \dot u (a)  v(a) - u(a) \dot v  (a) }  ~,~~
P_b= C^{-1}   \frac{1}{v(b)}
 \frac{\dot{f}(b)  v(b) - {\dot{v}(b)}  f(b)}{\dot{u}(b)v(b)-u(b)\dot{v}(b)}   \\
p(t) &=&   - C^{-1} {d\over dt}  \Big ( \frac{1}{v(t)}
 \frac{\dot{f}(t)  v(t) - {\dot{v}(t)}  f(t)}{\dot{u}(t)v(t)-u(t)\dot{v}(t)}   \Big) \frac{1}{v(t)}
\end{eqnarray*}
respectively.
Now, if $\hat \theta_n = \sum_{i=1}^n \omega_i Y_{t_i}$ denotes an unbiased linear estimate
in model \eqref{eq:model} with vectors $ \omega_i  \in \er^m$, we can represent this estimator as
$$
\hat \theta_n = \int_a^b Y_t \,  \hat \mu_n (dt) ,
$$
in the continuous time model \eqref{mod1cont}, where $ \hat \mu_n $ is a discrete signed vector valued measure with ``masses''
$\omega_i$ at the points $t_i$. Consequently, we obtain from  Theorem \ref{thm1} that
$$ C^{-1} =
\mbox{Var}(\hat{\theta}_{\rm BLUE} )  \leq  \mbox{Var}(\hat{\theta}_n) ,
$$
(in the Loewner ordering). In other words, $C^{-1}$ is a lower bound for any linear estimator in the linear regression model \eqref{eq:model}.

\end{itemize}
 }
 \end{remark}

\section{Optimal estimators and designs for finite sample size}
\label{sec3}
\def\theequation{3.\arabic{equation}}
\setcounter{equation}{0}

We have determined the BLUE and corresponding minimal variance/covariance matrix in the continuous time model
\eqref{mod1cont}. In the present section we now explain how the particular representation of the BLUE as a stochastic integral can be
used  to derive efficient estimators and corresponding optimal designs in the original model \eqref{eq:model},
which are practically not distinguishable from the BLUE in model \eqref{eq:model} based on an optimal design.
 Our approach is based on a comparison
of the mean squared error of the difference between the best linear unbiased estimator derived in Theorem \ref{thm1} and a discrete approximation
 of the stochastic integral in \eqref{asblue}. For the sake of a clear representation, we discuss the one-dimensional case first.

\subsection{One-parameter models} 

Consider the estimator  $\hat \theta_{\rm BLUE} $ defined by \eqref{asblue}  for  the continuous time model  \eqref{mod1cont} with $m=1$
and define an estimator $\hat \theta_n $ in the original regression model  by an approximation of the stochastic integral, that is
\begin{equation}\label{estimator}
\hat{\theta}_n = C^{-1} \Big\{ \sum_{i=2}^n \omega_i \dot{f}(t_{i-1}) (Y_{t_i}-Y_{t_{i-1}}) + \frac{f(a)}{a} Y_a \Big\}.
\end{equation}
Here  $a = t_1 < t_2 < \ldots < t_{n-1} < t_n=b $ are $n$ design points in the interval $[a,b]$ and ${\omega_2, \ldots, \omega_n}$ are
corresponding (not necessarily positive) weights.
 Obviously, the estimator depends on the weights $\omega_i$ only through the quantities $\mu_i=\omega_i \dot{f}(t_{i-1})$ and therefore we use the notation
\begin{equation}\label{estimator1}
 \hat{\theta}_n = C^{-1} \Big\{ \sum_{i=2}^n \mu_i   (Y_{t_i}-Y_{t_{i-1}}) + \frac{f(a)}{a} Y_a \Big\},
\end{equation}
 in the following discussion.
 We will determine optimal  weights ${\mu^*_2, \ldots, \mu_n^*}$ and design points $t_2^*, \ldots, t_{n-1}^*$
 minimizing the mean squared error $\mathbb{E}[(\hat{\theta}_{\rm BLUE}  - \hat{\theta}_n)^2]$
 between the estimators $\hat{\theta}_{\rm BLUE}$ and $ \hat{\theta}_n$.
Our first result provides an explicit expression for this quantity. The proof is omitted because we prove a more general  result later in the multi-parameter case (see Section \ref{a3}).

 \begin{lemma} 
Consider the continuous time model \eqref{mod1cont} in  the one-dimensional case. If the assumptions of Theorem \ref{thm1} are satisfied, then
\begin{align}
\mathbb{E}_\theta[(\hat{\theta}_{\rm BLUE}  - \hat{\theta}_n)^2] = &C^{-1} \Big\{ \sum_{i=2}^n \int_{t_{i-1}}^{t_i} \big[ \dot{f}(s) - \mu_i) \big]^2 \,ds \nonumber \\
&+ \theta^2 \Big( \sum_{i=2}^n \int_{t_{i-1}}^{t_i} \big[ \dot{f}(s) - \mu_i \big]\dot{f}(s) \,ds \Big)^2 \Big \} C^{-1} .
 \label{criterion-one}
\end{align}
\end{lemma}

In order to find ``good''weights for the linear estimator $\hat \theta_n$ in \eqref{estimator} we propose to consider only estimators with weights
${\mu_2, \ldots, \mu_n}$ such that the second term in \eqref{criterion-one} vanishes, that is
\begin{align}
&\sum_{i=2}^n \int_{t_{i-1}}^{t_i} \big[ \dot{f}(s) - \mu_i \big] \dot{f}(s) \,ds = 0.
\label{unbiased-one}
\end{align}
It is easy to see that this condition is  equivalent to the property that the estimator $\hat{\theta}_n$ in \eqref{estimator} is also unbiased, that is
 $\mathbb{E} [ \hat{\theta}_n ] = \theta$, or equivalently
 \begin{align}\label{new}
&\sum_{i=2}^n \mu_i ( {f}(t_i) - {f} (t_{i-1})) = \int^b_a [\dot{f}(s)]^2   ds.
\end{align}
 The following result describes the weights  minimizing $\mathbb{E}[(\hat{\theta}_{\rm BLUE}  - \hat{\theta}_n)^2] $
under the constraint \eqref{unbiased-one}.

 \begin{lemma}\label{optweightone}
Consider the continuous time model \eqref{mod1cont} in the one-dimensional case. If the assumptions of Theorem \ref{thm1} are satisfied, then
 the optimal weights minimizing $\mathbb{E}[(\hat{\theta}_{\rm BLUE}  - \hat{\theta}_n)^2] $
in the class of all unbiased linear estimators of the form \eqref{estimator} are given by
\begin{equation}\label{omega-one2}
\mu_i^* =  \kappa  (t_1,\ldots ,t_n) \frac{ f(t_i) - f(t_{i-1}) }{t_i-t_{i-1}} \, ,
\end{equation}
where \begin{equation*} 
\kappa  (t_1,\ldots ,t_n) =  \frac{\int_a^b [\dot{f}(s)]^2 \,ds}{\sum_{j=2}^n [f(t_j) - f(t_{j-1})]^2 / (t_j-t_{j-1})}\,.
\end{equation*}
\end{lemma}

{\bf Proof of Lemma \ref{optweightone}.}  Under the condition \eqref{unbiased-one} the mean squared error simplifies to
\begin{align*}
\mathbb{E}_\theta[(\hat{\theta}_{\rm BLUE}  - \hat{\theta}_n)^2] &= C^{-1} \Big \{ \sum_{i=2}^n \int_{t_{i-1}}^{t_i} \big[ \dot{f}(s) -  \mu_i \big]^2 \,ds \Big \}   C^{-1} \\
&= C^{-1} \Big\{ - \int_a^b [\dot{f}(s)]^2 \,ds + \sum_{i=2}^n \mu_i^2 (t_i - t_{i-1}) \Big\} C^{-1} .
\end{align*}
Using Lagrangian multiplies to minimize this expression subject to the constraint \eqref{new} yields
\begin{align*}
&  \mu_i = \frac{\lambda [f(t_i) - f(t_{i-1})]}{2  (t_i-t_{i-1})}~,  \quad  i =2, \ldots, n ,
\end{align*}
where $\lambda$ denotes the Lagrangian multiplier. Substituting this into \eqref{unbiased-one} gives
\begin{align*}
&\lambda/2 = \frac{\int_a^b [\dot{f}(s)]^2 \,ds}{\sum_{i=2}^n [f(t_i) - f(t_{i-1})]^2 / (t_i-t_{i-1})} = \kappa  (t_1,\ldots ,t_n) .
\end{align*}
Therefore, the optimal weights are given by \eqref{omega-one2}. \hfill $\Box$.

\bigskip

Inserting these weights in the mean squared error
 gives the function
 $$
 \mathbb{E}_\theta[(\hat \theta_{\rm BLUE}  - \hat \theta_n)^2] ~=~
C^{-1} \Big \{ \Big ( \int^b_a [ \dot{f} (s)] ^2ds  \Big)^2 \Big\{ \sum^n_{i=2} \frac {(f(t_i)-f(t_{i-1}))^2}{t_i-t_{i-1}} \Big\}^{-1} - \int^b_a [ \dot{f} (s)] ^2ds \Big \} C^{-1},
$$
which finally has to be minimized by the choice of the design points $t_2,\ldots,t_{n-1}$.
 Because we discuss the one-parameter case in this section and the matrix $C$ does not depend on $t_2, \ldots, t_n$, this optimization corresponds to the minimization of
 \be\label{crit1}
 \Phi (t_1,\dots,t_n) =
  \Big(  \int^b_a [ \dot{f} (s)]^2 ds \Big)
 \Big \{ \sum^n_{i=2} \frac {(f(t_i)-f(t_{i-1}))^2}{t_i - t_{i-1}} \Big \}^{-1}  - 1    .
\ee

\begin{remark}  \label{rem43}
{\rm
Let
\begin{eqnarray*}
\mbox{eff} (t_2,\ldots, t_{n-1}) &=&  \frac {\mbox{Var}_\theta (\hat \theta_{\rm BLUE})}{\mbox{Var}_\theta(\hat \theta_n )]} = \frac {C^{-1}} {C^{-1}
 \int_a^b [\dot{f}(s)]^2 ds \, \Phi (t_1,\ldots,t_n) C^{-1}+C^{-1}}  \\
&=& \left( 1+ \frac {\Phi (t_1,\ldots,t_n)}{1+ \frac {f^2(a)}{a} / \int^b_a [\dot{f}(s)]^2ds} \right) ^{-1} ,
\end{eqnarray*}
denote the efficiency of an estimator $\hat \theta_n$ defined by \eqref{estimator} with optimal weights.
Note that from the proof of Lemma \ref{optweightone} it follows that the function $\Phi$ is non-negative for all $t_1,\dots,t_n$.
Consequently, minimizing $\Phi$  with respect to the design points means that $t_1=a<t_2< \ldots < t_{n-1}<t_n=b$ have to be determined such that
$$
\sum^n_{i=2} \frac {(f(t_i) - f(t_{i-1}))^2}{t_i - t_{i-1}} ,
$$
approximates the integral $\int^b_a [\dot{f}(s)]^2 ds$ most precisely (this produces an efficiency close to $1$).
Now, if $f$ is sufficiently smooth, we have for any $\xi_i \in [t_{i-1},t_i]$
$$
\Big |  \frac {( f(t_i) - f(t_{i-1}))^2  }{t_i - t_{i-1}}  -  [ \dot {f} (\xi_i)]^2 (t_i - t_{i-1})  \Big| ~\le ~ G,
$$
for all $i=2,\ldots,n$, where
\begin{eqnarray*}
 G:=  2 \max_{\xi \in [a,b] } | f^{\prime } (\xi )|  \max_{\xi \in [a,b] }  | f^{\prime\prime} (\xi ) |\cdot \max_{i=2,\ldots,n}|t_i - t_{i-1}|^2 .
 \end{eqnarray*}
 This gives
 $$
 0 \leq A(t_1,\ldots,t_n):= \int^b_a \dot f^2 (t) dt - \sum^n_{i=2} \frac {(f(t_i)-f(t_{i-1}))^2}{t_i - t_{i-1}} \leq (n-1) G.
 $$
 As the function $\Phi$ has the representation
 $$
 \Phi (t_1,\ldots,t_n)= \frac {A (t_1,\ldots,t_n)}{\int^b_a \dot f^2 (s) ds - A(t_1,\ldots,t_n)}
 $$
 it follows that (note that the expression on the right-hand side is increasing with $A(t_1,\ldots,t_n)$)
\begin{equation} \label{bd}
\Phi (t_1,\ldots,t_n) \leq \frac { {(n-1) \cdot \max_{i=2,\ldots,n} |t_i - t_{i-1}|^2}}
 {\displaystyle {H(f) + (n-1) \cdot \max_{i=2,\ldots,n}    |t_i - t_{i-1}|^2}},
\end{equation}
where
$$
H(f) = \frac {\int^b_a \dot f^2 (s) ds}
{\displaystyle{2 \max_ {\xi \in [a,b]}  |\dot f(\xi)| \max_{\xi \in [a,b]} | \ddot f(s) |}}.
$$
This shows that for most models a substantial improvement of the approximation  by the choice of $t_2,\ldots,t_n$ can only be achieved if the sample
size is   small. For moderate or large sample sizes  one could use the points $ u_i = a + {i-1 \over n-1} (b-a) $,
 which gives already the estimate
$$
\Phi (u_1, \ldots,u_n)  \leq  \frac {1}{1+(n-1)H(f)} =  O \bigl(\frac {1}{n}\bigr)
$$
(note that we consider worst case scenarios to obtain these estimates).
Consequently, in many cases
the design points can be chosen in an equidistant way, because the
choice of the points $t_2,\ldots,t_{n-1}$ is irrelevant from a practical point of view,  provided that
the weights of the estimator $\hat \theta_n$ are already
chosen in an optimal way.
}
\end{remark}

\begin{example} \label{exquad} {\rm
Consider the quadratic regression model $Y_t=\theta t^2+\varepsilon_t$, where $t \in [a,b]$. Then $f(t)=t^2, \ \dot{f}(t) = 2t$, and the function $\Phi$ in \eqref{crit1} reduces to 
$$
\Phi (t_1,\ldots,t_n)= \frac {4(b^3-a^3)}{3} \Big \{ \sum^n_{i=2} (t_i + t_{i-1})^2 (t_i - t_{i-1}) \Big \}^{-1} - 1.
$$
It follows by a straightforward computation that the optimal points are given by
\begin{equation} \label{optquad}
t^*_i = a + \frac {i-1}{n-1} (b-a) \, ; \qquad i=1,\ldots,n,
\end{equation}
while the corresponding minimal value is 
$$
\Phi (t^*_1, \ldots, t^*_n)= \frac{(a-b)^3}{4(n-1)^2(a^3-b^3)-(a-b)^3} \qquad (n \geq 2).
$$
Note that  this term is of order $O({1\over n^2})$.  Remark \ref{rem43} gives the bound
$$
\Phi (t^*_1, \ldots, t^*_n) \leq \frac {1}{1+ \frac {b^3-a^3}{2b}(n-1)} = O\big({1\over n} \big)~,
$$
which shows that \eqref{bd} is not necessarily sharp.
For the efficiency we obtain 
$$
\mbox{eff}(t^*_1,\ldots,t^*_n)=1- \frac{4(a-b)^3(a^3-b^3)}{3a^3(a-b)^3+4(n-1)^2(a^3-b^3)(a-b)^3},
$$
which is of order $1- O({1\over n^2})$.
On the other hand, if $f(t)=t^3$ the function $\Phi$ is given by
\begin{align*}
\Phi (t_1, \ldots, t_n) &= \frac {9}{5} (b^5-a^5) \Big \{ \sum^n_{i=2} (t_i - t_{i-1}) (t_i^2 + t_i t_{i-1} + t^2_{i-1})^2 \Big \}^{-1}-1 \\
&=\frac{(a-b)^2[5(n-1)^2(a^3-b^3)-(a-b)^3]}{9(n-1)^4(a^5-b^5)-(a-b)^2[5(n-1)^2(a^3-b^3)-(a-b)^3]}
\end{align*}
and optimal points have to be found numerically. However, we can evaluate the efficiency of the uniform design in \eqref{optquad}, which is given by
$$
\mbox{eff} (t^*_1, \ldots, t^*_n) = 1-\frac{9(b^5-a^5) (a-b)^2[5(n-1)^2(a^3-b^3)-(a-b)^3]}{9(9b^5-4a^5)(a^5-b^5)(n-1)^4+5a^5 (a-b)^2[5(n-1)^2(a^3-b^3)-(a-b)^3]}
$$
$ (n \geq 2)$ and also of order $1- O({1\over n^2})$.
Thus, although the uniform design is not optimal, its efficiency (with respect to the continuous case) is extremely high.}
\end{example}

\subsection{Multi-parameter models} 

In this section we derive corresponding results for the multi-parameter case. If $m \geq 1$  we propose a linear estimator with matrix weights as an analogue of \eqref{estimator}, that is
\begin{eqnarray}\label{estimatormul}
\hat{\theta}_n &=& C^{-1} \Big\{ \sum_{i=2}^n \Omega_i \dot{f}(t_{i-1}) (Y_{t_i}-Y_{t_{i-1}}) + \frac{f(a)}{a} Y_a \Big\} \\ \nonumber
&=&  C^{-1} \Big\{ \sum_{i=2}^n \mu_i   (Y_{t_i}-Y_{t_{i-1}}) + \frac{f(a)}{a} Y_a \Big\},
\end{eqnarray}
where $C^{-1}$ is given in \eqref{cmatrix1}, $\Omega_2, \ldots, \Omega_n$ are $m \times m$ matrices and $\mu_2=\Omega_2 \dot{f}(t_i), \ldots, \mu_n = \Omega_n \dot{f} (t_{n-1})$ are $m$-dimensional vectors, which have to be chosen in a reasonable way. For this purpose we
derive a representation of the mean squared error between the best linear estimate in the continuous time model and
its discrete approximation in the multi-parameter case first. The proof can be found in   Appendix \ref{a3}.

 \begin{lemma} \label{criterion-multi-dimension}
Consider the continuous time model \eqref{mod1cont}. If the assumptions of Theorem \ref{thm1} are satisfied, then
\begin{align}
&\mathbb{E}_\theta[(\hat{\theta}_{\rm BLUE}  - \hat{\theta}_n)(\hat{\theta}_{\rm BLUE}  - \hat{\theta}_n)^T] = C^{-1} \Big\{ \sum_{i=2}^n \int_{t_{i-1}}^{t_i} \big[ \dot{f}(s) -  \mu_i \big] \big[ \dot{f}  (s) - \mu_i   \big]^T \,ds \nonumber \\
&+ \sum_{i=2}^n \int_{t_{i-1}}^{t_i} \big[ \dot{f}(s) - \mu_i \big]\dot{f}^T(s) \,ds \, \theta \, \theta^T
 \sum_{j=2}^n \int_{t_{j-1}}^{t_j}  \dot{f}(s) \big[ \dot{f} (s) - \mu_j \big]^T \,ds \Big\} C^{-1}.
 \label{criterion-multi}
\end{align}
\end{lemma}

\bigskip

In the following we choose optimal vectors (or equivalently matrices $\Omega_i$) $\mu_i=\Omega_i \dot{f}(t_{i-1})$ and design points $t_i$, such that the linear estimate \eqref{estimatormul} is unbiased and    the mean squared error matrix in \eqref{criterion-multi} ``becomes small''.
An  alternative criterion is to replace the mean squared error $\mathbb{E}_\theta[(\hat \theta_{\rm BLUE}  - \hat \theta_n)(\hat \theta_{\rm BLUE}  - \hat \theta_n)^T]$  by the mean squared error
$$\mathbb{E}_\theta [(\hat{\theta}_n - \theta)(\hat{\theta}_n - \theta)^T]$$  between the   estimate $\hat \theta_n$ defined in \eqref{estimatormul} and the ``true'' vector of parameters.
The following result shows that both optimization problems will yield the same solution in the class of all unbiased estimators. The proof can be found in   Appendix \ref{a4}.

\begin{theorem}\label{equivalence}
The estimator $\hat{\theta}_n$ defined in \eqref{estimator} is unbiased if and only if the identity
\begin{equation} \label{unbiasmul}
\int_a^b \dot{f}(s) \dot{f}^T(s) \,ds = \sum_{i=2}^n \mu_i \int_{t_{i-1}}^{t_i} \dot{f}^T(s) \,ds = \sum^n_{i=2}   \mu_i(f(t_i)-f(t_{i-1}))^T ,
\end{equation}
is satisfied.
Moreover, for any linear unbiased estimator of the form $\tilde{\theta}_n = \int_a^b g(s) dY_s $ we have
\begin{equation*}
\mathbb{E}_\theta [(\tilde{\theta}_n - \theta)(\tilde{\theta}_n - \theta)^T] = \mathbb{E}_\theta [(\tilde{\theta}_n - \hat{\theta}_{\rm BLUE} )(\tilde{\theta}_n - \hat{\theta}_{\rm BLUE} )^T] + C^{-1}.
\end{equation*}
\end{theorem}

\bigskip

In order to describe a solution in terms of optimal ``weights'' $\mu^*_i$ and design points $t^*_i$ we recall that the condition of unbiasedness of the estimate $\hat \theta_n$ in \eqref{estimatormul} is given by \eqref{unbiasmul} and
introduce the notation
\begin{align}
& \beta^{(i)} = [f(t_i) - f(t_{i-1})] / \sqrt{t_i-t_{i-1}},  \label{h1} \\
& \gamma^{(i)} = \mu_i \sqrt{t_i-t_{i-1}}.\nonumber
\end{align}
It   follows from Lemma \ref{criterion-multi-dimension} that for an unbiased estimate $\hat \theta_n$
the mean squared error has the representation
\be \label{crit}
 \mathbb{E}_\theta \big  [( \hat \theta_{\rm BLUE}  - \hat \theta_n)^T (\hat \theta_{\rm BLUE}  - \hat \theta_n)\big ]
  =
 -C^{-1} M C^{-1}   + \sum_{i=2}^nC^{-1}  \gamma^{(i)} \gamma^{(i)^T} C^{-1}   ,
\ee
which has to be ``minimized'' subject to the constraint
\be \label{crit0}
M=(m_{\ell,k})_{\ell,k}^m = \int^b_a \dot{f} (s) \dot{f}^T(s)ds=  \sum_{i=2}^n \gamma^{(i)} \beta^{(i)^T}.
\ee
The following result shows that a minimization with respect to  the weights $\mu_i$ (or equivalently $\gamma_i$) can actually be carried out with respect to
the Loewner ordering.

\begin{theorem} \label{thm3}
Assume that the assumptions of Theorem \ref{thm1} are satisfied and that the matrix
$$
B = \sum^n_{i=2} \frac {[f(t_i)-f(t_{i-1})][f(t_i) - f(t_{i-1})]^T  }{t_i - t_{i-1}} ,
$$
is non-singular. Let  $\mu^*_2, \ldots, \mu^*_n$ denote $m \times 1$ vectors satisfying the equations
\be \label{eq1}
\mu^*_i = M B^{-1} \frac {f(t_i) - f(t_{i-1})}{t_i - t_{i-1}} \qquad i=2,\ldots,n,
\ee
then $\mu^*_2, \ldots, \mu^*_n$ are optimal (vector) weights minimizing $ \mathbb{E}_\theta[(\hat{\theta}_{\rm BLUE}
 - \hat{\theta}_n)(\hat{\theta}_{\rm BLUE} - \hat{\theta}_n)^T] $
 with respect to the Loewner ordering
among all unbiased estimators of the form \eqref{estimatormul}.
\end{theorem}

\textbf{Proof of Theorem \ref{thm3}.}  Let $A$ denote a positive definite $m\times m$ matrix and consider the problem of  minimizing
 the linear criterion
 $$\mbox{tr} \left\{A\hspace{0.2cm}  \mathbb{E}_\theta[(\hat{\theta}_{\rm BLUE}  - \hat{\theta}_n)(\hat{\theta}_{\rm BLUE}
 - \hat{\theta}_n)^T]  \right\}$$
  subject to the constraint \eqref{crit0}. Observing \eqref{crit} this yields the Lagrange function
\begin{equation*}
- \mbox{tr}  \{ A C^{-1} M C^{-1} \} + \sum^n_{i=2} (C^{-1} \gamma^{(i)})^T A (C^{-1}\gamma^{(i)}) - \sum^m_{k,\ell=1} \lambda_{k,\ell} \Big(m_{k,\ell} - \sum^n_{i=2} \gamma^{(i)}_k \beta^{(i)}_\ell \Big ),
\end{equation*}
where $C=(c_{k,\ell})^m_{k,\ell =1}, \ \gamma^{(i)} = (\gamma^{(i)}_1, \ldots, \gamma^{(i)}_m)^T, \
\beta^{(i)} = (\beta^{(i)}_1, \ldots, \beta^{(i)}_m)^T$ and $\Lambda = (\lambda_{k,\ell})^m_{k,\ell =1}$ is a matrix of Lagrange multipliers. This function is obviously convex with respect to $\gamma^{(2)},\ldots,\gamma^{(n)}$. Therefore, taking derivatives with respect to $\gamma^{(i)}_j$ yields as necessary and  sufficient  for the extremum
$$
\sum_{p=1}^m c^{p,j} \sum_{\ell=1}^m a_{p,\ell} \sum_{k=1}^m c^{\ell,k} \gamma_k^{(i)}  + \sum_{p=1}^m \sum_{k=1}^m c^{p,k} \gamma_k^{(i)} \sum_{\ell=1}^m a_{p,\ell} c^{\ell,j} + \sum^m_{\ell=1} \lambda_{j,\ell} \beta^{(i)}_\ell = 0 \qquad j=1,\ldots,k,
$$
where $A=(a_{\ell,k})^m_{\ell,k=1}$ and $C^{-1}=(c^{\ell,k})^m_{\ell,k=1}$ is the inverse of the matrix $C$ defined in \eqref{asblue}. Rewriting this system of linear equations in matrix form gives
$$
C^{-1} A C^{-1} \gamma^{(i)} + C^{-1} A^T C^{-1} \gamma^{(i)} + \Lambda \beta^{(i)} = 0  \qquad i=2,\ldots,n,
$$
or equivalently
$$
 C^{-1} (A + A^T) C^{-1} \gamma^{(i)} = -\Lambda \beta^{(i)} \qquad i=2,\ldots,n.
$$
Substituting this expression in \eqref{crit0} and using the non-singularity of the matrices $C$ and $B$ yields for the matrix of Lagrangian multipliers
$$
\Lambda = - C^{-1} (A + A^T) C^{-1} MB^{-1},
$$
which finally gives
$$
\gamma^{(i)} = M B^{-1} \beta^{(i)} \qquad i=2,\ldots,n.
$$
Observing the notations in \eqref{h1} shows  that the optimal vector weights are given by  \eqref{eq1}.
 Thus  the optimal weights in \eqref{eq1}  do not depend on the matrix $A$ and
 provide the solution  for all linear optimality criteria. Consequently, using the matrices
$A= vv^T + \varepsilon I_m$ with $v \in \er^m$, and considering the limit as $\varepsilon \to 0$, shows that the weights defined in \eqref{eq1} minimize $ \mathbb{E}_\theta[(\hat{\theta}_{\rm BLUE}
 - \hat{\theta}_n)(\hat{\theta}_{\rm BLUE} - \hat{\theta}_n)^T] $  with respect to the Loewner ordering.

\hfill $\Box$

\bigskip

\begin{remark} 
{\rm
If the matrix $B$ in Theorem \ref{thm3} is singular, the optimal vectors are not uniquely determined and we propose to replace the inverse $B$ by its Moore-Penrose inverse.}
\end{remark}

Note that for fixed design points $t_1, \ldots , t_n $ Theorem \ref{thm3} yields
universally optimal weights $\mu^*_2,\ldots,\mu^*_n$ (with respect to the Loewner ordering)
for  estimators of the form \eqref{estimatormul} satisfying \eqref{unbiasmul}. On the other hand, a further optimization with respect to the Loewner ordering with respect to the choice of the points $t_1,\ldots,t_n$ is not possible, and we have to apply a real valued optimality criterion for this purpose. More precisely, let $\hat \theta^*_n$ denote the estimator of the form \eqref{estimatormul} with optimal weights $\gamma^{*(i)}= \mu_i^* \sqrt{t_i - t_{i-1}}$ given by \eqref{eq1}, then we choose $t_1,\ldots,t_n$, such that
\begin{eqnarray*}
&& \mbox{tr} \big( \mathbb{E}_\theta \big  [( \hat \theta_{\rm BLUE}  - \hat \theta^*_n)^T (\hat \theta_{\rm BLUE}  - \hat \theta^*_n)\big ]\big)
  =\mbox{tr} \Big \{ -C^{-1} M C^{-1}   + \sum_{i=2}^nC^{-1}  \gamma^{^*(i)} \gamma^{^*(i)^T} C^{-1}   \Big\}  \\
&& ~~~~~~~~~~~~~~~~~~~~= {\rm tr} \Bigl \{ - C^{-1} M C^{-1} + C^{-1}M \Bigl ( \sum^n_{i=2} \frac {(f(t_i)-f(t_{i-1})(f(t_i)-f(t_{i-1}))^T}{t_i - t_{i-1}} \Bigr)^{-1} MC^{-1} \Bigr \}
\end{eqnarray*}
is minimal. The performance of this method will be illustrated in the following section.

\section{Some numerical examples}
\label{sec4}
\def\theequation{4.\arabic{equation}}
\setcounter{equation}{0}

In this section we illustrate our new methodology using several model and covariance kernel examples. Note that (under smoothness assumptions) our
approach allows us to calculate a lower bound for the trace (or any other monotone functional)
of the variance of any (unbiased) linear estimator
for the parameter vector $\theta$  in model \eqref{eq:model} [see Remark \ref{rem1}(c)]. Therefore
we evaluate the quality of an  estimator (with corresponding design), say $\hat \theta$,   by the efficiency
\begin{equation*} 
\text{eff}(\hat{\theta}) ~= ~\frac{\mbox{tr} \{\mbox{Var}_\theta (\hat{\theta}_{BLUE})\}}{\mbox{tr} \{\mbox{Var}_\theta
(\hat{\theta})\}}~ =~ \frac{\mbox{tr} ( C^{-1}) }{\mbox{tr} \{\mbox{Var}_\theta(\hat{\theta})\}}  ,
\end{equation*}
Throughout this section the estimator defined by \eqref{estimator1} and Lemma \ref{optweightone} in the case of $m=1$ and by \eqref{estimatormul} and Theorem \ref{thm3} for $m>1$, will be denoted by $\hat{\theta}_n^*$.
As before the univariate and multivariate cases are studied separately.

\subsection{One-parameter models} 

Consider  model  \eqref{eq:model}  with $m=1$ and $n=5$  observations in the interval $[a, b] = [1, 2]$,
where the regression function is given by  $f(t)=t^2$, $t^2-0.5$ and  $t^4$ with kernel $k(s,t)=s \wedge t$.
The discussion in Example \ref{exquad}
indicates that equally spaced design points provide already an efficient allocation for the new estimator
$\hat \theta_n^*$. Consequently,   we compare the estimator
$\hat{\theta}_{{\rm DPZ},n}$ (with a corresponding optimal design)
proposed in  Section 2.5 of \cite{detpepzhi2015} with the
 BLUE  and also with  the estimator  defined by  \eqref{estimator1}  and Lemma \ref{optweightone} based
on a uniform design.
The latter  two estimators are denoted by  $\hat{\theta}_{{\rm BLUE},n}^{\rm uni}$  and   $\hat{\theta}_{n}^{* \rm uni}$, respectively,  and we consider
a uniform design with $n=5$  points.
The corresponding efficiencies are displayed in Table \ref{tab1}.

\begin{table}[htp!]
\caption{\label{tab1}
\emph{Efficiencies (in percent)  of various estimators in the univariate linear regression model for $n=5$ observations on the interval $[1,2]$.
$\hat{\theta}_{{\rm BLUE},n}^{\rm uni} $ is the BLUE
based on a uniform design, $\hat{\theta}_{n}^{* \rm uni}$  is the estimator defined by \eqref{estimator1} and Lemma \ref{optweightone} based on a uniform design
and $\hat{\theta}_{{\rm DPZ},n}$  (with a corresponding   design)
proposed in  \cite{detpepzhi2015}.  }}
\vspace{0.5cm}
\centering
\begin{tabular}{|c|c|c|c|}

\hline

\cline{2-4}
$f(t) $& $t^2$ & $t^2-0.5$ & $t^4$  \\
\hline
$\hat{\theta}_{{\rm BLUE},n}^{\rm uni} $ & 99.798 & 99.783 & 98.416  \\
\hline
$\hat{\theta}_{n}^{* \rm uni}$  & 99.798 & 99.783 & 98.416  \\
\hline
$\hat{\theta}_{{\rm DPZ},n}$   & 99.582 & 99.346 & 92.662  \\
\hline

\end{tabular}
\end{table}

We observe that both $\hat{\theta}_{{\rm BLUE},n}^{\rm uni}$  and   $\hat{\theta}_{n}^{* \rm uni}$ have very good efficiencies and therefore we did not determine the optimal allocations for the two estimators. A comparison between both estimators shows that $\hat{\theta}_{{\rm BLUE},n}^{\rm uni}$ and $\hat{\theta}_{n}^{* \rm uni}$      are practically not distinguishable. In   all the cases considered, the efficiencies do not differ in the first 5 decimals. For example,
for the function $f(t)=t^2-0.5$ we have
$$
\text{eff}(\hat{\theta}_{{\rm BLUE},n}^{\rm uni} )  = 0.99782609 ~,~\text{eff}( \hat{\theta}_{n}^{* \rm uni})=  0.99782596 ~.
$$
The investigation of other one-dimensional examples showed a similar picture and details are omitted for the sake of brevity. Therefore, the new estimator $\hat{\theta}_{n}^{* \rm}$ with a uniform design is not only highly efficient (even for small values of $n$),
but most importantly, it is very close to the best achievable.   The comparison with
the estimator $\hat{\theta}_{{\rm DPZ},n}$ proposed in  \cite{detpepzhi2015}  shows that
the new approach still provides an improvement of
an estimator which has efficiencies already above $90\%$, with the difference of efficiencies being small for $f(t) = t^2, t^2-0.5$ and large for $f(t) = t^4$.

\subsection{Models with $m>1$ parameters} 

We now compare the various estimators in the multi-parameter case. In particular, we consider two regression models given by
 \begin{eqnarray}
\label{mod1} Y_t &=&  (t, t^2, t^3)^T \theta  + \varepsilon_t, \quad t \in [a,b] \\
\label{mod2} Y_t &= &  \left(\sin t, \cos t, \sin 2t, \cos 2t \right)^T \theta  + \varepsilon_t, \quad t \in [a,b].
\end{eqnarray}
For each one of these models we study two cases of the covariance kernel of the error process in model \eqref{eq:model}, namely     $K(t,t') = \min \{ t, t' \}$ and $K(t, t') = \exp \{ - \lambda |t-t'| \}$. The sample size is again $n=5$ and the design space is the interval $[1,2]$.

 It turns out that for these models and the particularly small sample size the uniform design does not yield similar high efficiencies as in the case $m=1$ discussed in the previous section. For this reason we also calculate the corresponding optimal designs for the BLUE   $\hat{\theta}_{{\rm BLUE},n}  $
  and the   estimator  $\hat{\theta}_{n}^*$  proposed in  this paper  [see \eqref{estimatormul} and Theorem  \ref{thm3}]
 using the Particle swarm optimization (PSO) algorithm [see for example  \cite{Clerc2006}  or \cite{wongmix} among others].  \\
If the error process is a Brownian motion, the optimal design of $\hat{\theta}_{n}^*$ is obtained by applying the PSO algorithm on the trace of the mean squared error $\mathbb{E}_\theta[(\hat{\theta}_{\rm BLUE} - \hat{\theta}_n)(\hat{\theta}_{\rm BLUE} - \hat{\theta}_n)^T]$ given in \eqref{crit} (or equivalently on the trace of $\mathbb{E}_\theta[(\hat{\theta}_n - \theta)(\hat{\theta}_n - \theta)^T]$), using the optimal weights $\mu_i^*$, $i=2, \ldots, n$, given in Theorem \ref{thm3}. In the case of the exponential kernel $K(t, t') = \exp \{ - \lambda |t-t'| \}$ we follow the same procedure as before but for the transformed continuous time model given in \eqref{transformed-model}. The optimal design for the initial model with the exponential covariance kernel can then be obtained by the transformation $\tilde{t}=q(t)$ applied on each one of the optimal design points the algorithm will yield (see Remark \ref{rem1}(b)). Minimizing (using the PSO method) the trace of Var$(\hat{\theta}_{\rm WLSE})$ given in \eqref{eq:var-wls} for the corresponding variance/covariance matrix $\Sigma=\left( K(t_i,t_j) \right)_{i,j=1,\ldots,n}$ of the error process gives the optimal design for $\hat{\theta}_{\rm BLUE,n}$. \\
For the model and covariance kernel examples under consideration, the optimal designs for the estimators $\hat{\theta}_{BLUE,n}$ and $\hat{\theta}_n^*$ are presented in Table \ref{tab2}. The corresponding designs for the estimator $\hat{\theta}_{\rm DPZ}$ are chosen as described in \cite{detpepzhi2015}.
We observe that regardless of the model and the covariance kernel, the optimal designs for the estimators $\hat{\theta}_{\rm BLUE,n}$ and $\hat{\theta}_n^*$ are very similar. Furthermore, for the specific examples, the choice of covariance kernel does not affect the optimal design since for a given estimator, the two kernels yield the same design (up to 2 d.p.) for both models. In particular, the optimal designs are always supported at both end-points of the design space. For model \eqref{mod1}, although the uniform design is not optimal, the middle points of the optimal design are somewhat spread in the interval $(1,2)$, whereas in the case of model \eqref{mod2}, more points are allocated closer to the lower bound $t=1$ of the design space.
\begin{table}[!htbp!]
\caption{\label{tab2} \emph{Optimal five-point designs  in the interval $[1,2]$ for
the estimators $\hat{\theta}_{{\rm BLUE},n}$ and  $\hat{\theta}_{n}^*$ for models
\eqref{mod1} and \eqref{mod2} with two covariance kernels. }}
\vspace{0.5cm}
\centering
\begin{tabular}{|c|c|c|c|}
\hline
\multicolumn{2}{|c|}{} & \multicolumn{2}{|c|}{Optimal designs} \\
\hline
Model & Kernel & $\hat{\theta}_{{\rm BLUE},n}$  & $\hat{\theta}_{n}^*$  \\
\hline
\multirow{2}{*}{\eqref{mod1} } & $t \wedge t'$ & [1, 1.466, 1.680, 1.852, 2] & [1, 1.444, 1.668, 1.846, 2] \\
\cline{2-4}
& $\exp \{ - |t-t'| \}$ & [1, 1.474, 1.683, 1.852, 2] &  [1, 1.459, 1.674, 1.847, 2]\\
\hline
\multirow{2}{*}{\eqref{mod2}} & $t \wedge t'$ & [1, 1.111, 1.243, 1.800, 2] & [1, 1.120, 1.264, 1.802,2]  \\
\cline{2-4}
& $\exp \{ - |t-t'| \}$ & [1, 1.113, 1.245, 1.800, 2] & [1, 1.120, 1.263, 1.801, 2] \\
\hline
\end{tabular}
\end{table}

Table \ref{tab3} gives the efficiencies of the three estimators $\hat{\theta}_{{\rm BLUE},n}$, $\hat{\theta}_{n}^*$ and $\hat{\theta}_{{\rm DPZ},n}$ for the optimal design of each estimator (upper part) and the uniform design (lower part) with $n=5$ observations. For model \eqref{mod1} and any of the two covariance kernels, if the uniform design is used both  $\hat{\theta}_{{\rm BLUE},n}$ and $\hat{\theta}_{n}^*$ estimators are very efficient. The efficiencies of course increase when observations are taken according to the optimal instead of the uniform design   but remain below $90 \%$ when the four-dimensional model \eqref{mod2} is considered.
  \begin{table}[htp!]
\caption{\label{tab3}
\emph{Efficiencies (in percent) of the estimators $\hat{\theta}_{{\rm BLUE},n}$, $\hat{\theta}_{n}^*$ and $\hat{\theta}_{{\rm DPZ},n}$ for models \eqref{mod1} and \eqref{mod2} and for two covariance kernels of the error process. The design is the uniform or the optimal design for five observations}}
\vspace{0.5cm}
\centering
\begin{tabular}{|c|c|c|c|c|c|}
\hline
\multicolumn{3}{|c|}{} & \multicolumn{3}{|c|}{Efficiencies} \\
\hline
 & Model & Kernel & $\hat{\theta}_{{\rm BLUE},n}$ & $\hat{\theta}_{n}^*$ & $\hat{\theta}_{{\rm DPZ},n}$ \\
\hline
 \multirow{4}{*}{optimal design} & \multirow{2}{*}{\eqref{mod1}} & $t \wedge t'$ & 96.77 & 96.71 & 82.14 \\
\cline{3-6}
& & $\exp \{ - |t-t'| \}$  & 96.72  & 96.65 &  79.60   \\
\cline{2-6}
 & \eqref{mod2} & $t \wedge t'$ & 83.98 & 83.40 & 70.91 \\
\cline{3-6}
& & $\exp \{ - |t-t'| \}$  & 83.47  & 82.95 &  71.57   \\
\hline
\hline
\multirow{4}{*}{uniform design} & \multirow{2}{*}{\eqref{mod1}} & $t \wedge t'$ & 94.35 & 93.82 & 76.38 \\
\cline{3-6}
& & $\exp \{ - |t-t'| \}$  & 94.07  & 93.46 & 75.10   \\
\cline{2-6}
 & \eqref{mod2} & $t \wedge t'$ & 73.13 & 73.12 & 70.91 \\
\cline{3-6}
& & $\exp \{ - |t-t'| \}$  & 72.56  & 72.46 &  71.57  \\
\hline
\end{tabular}
\end{table}

We also observe that the estimator $\hat{\theta}_{n}^*$ proposed in this paper has substantially larger efficiencies than $\hat{\theta}_{{\rm DPZ},n}$ (always well below $90\%$) and thus the new approach provides a substantial improvement and is additionally much easier to implement for multi-parameter models than that introduced in \cite{detpepzhi2015}. Finally, the estimators $\hat{\theta}_{{\rm BLUE},n}$ and $\hat{\theta}_{n}^*$ have similar efficiencies regardless of the underlying design. We therefore conclude that the alternative approach proposed in this paper provides estimators with corresponding optimal designs for finite sample which are practically not distinguishable from the optimal estimator and corresponding design.

\section{Degenerate models}\label{secextra}

So far we have considered the continuous regression model \eqref{mod1cont} with a covariance kernel of the form
\eqref{eq:cov_tr0} satisfying $u(a) \neq 0$. If $u(a)=0$, then  the variance of the observation at $t=a$ is 0 and all formulas of Section \ref{sec2} and \ref{sec3} degenerate in this case. The estimator
$\hat{\theta}_{\rm BLUE}$ in the continuous time model and its discrete approximation \eqref{estimatormul} are
 not well defined and  the results of previous sections cannot be applied.
In this section, we indicate  how the methodology can be extended to the case $u(a)=0$. For the sake of brevity we
only consider the continuous time model with a Brownian motion as error process, since the transformation (\ref{trans})
which reduces any model with the covariance kernel \eqref{eq:cov_tr0} to the case of Brownian motion can still be applied.
Moreover, the construction of an estimator (with a corresponding design) from the solution for the continuous time model can be obtained by similar arguments as
presented in Section \ref{sec3}.

The main idea is to construct the BLUE $\hat{\theta}_{\rm BLUE}  $ in the continuous time  model \eqref{mod1cont} on the interval $[0,b]$
by a sequence of estimators $\hat{\theta}_{{\rm BLUE}  ,a}$ for the same model on the  interval $[a,b]$, where $a \to 0$.
For this purpose we
make the dependence of some quantities in the following discussion more explicit.
For example we write  $C_a$ for the matrix $C$
defined in \eqref{cmatrix} and so on.
We have to consider three different cases of degeneracy, which will be discussed below.

\subsection{Models with no intercept, that is  $1 \notin {\rm span}\{f_1, \ldots, f_m\} $}

By Lemma \ref{l:2 cases} in Section \ref{aux}, if $1 \notin {\rm span}\{f_1, \ldots, f_m\} $ then  the matrix
\bea 
M_a = \int^b_a \dot{f} (s) \dot{f}^T(s)ds
\eea
is non-singular for all $a \in [0,b)$.  In particular, $M_0^{-1}$ exists.
 Additionally, in this case, for any $a>0$ the inverse of the  matrix
 \begin{equation*} 
C_a = \int_a^b \dot{f}(t) \dot{f}^T(t) \,dt + \frac{f(a) f^T(a)}{a}= M_a+ \frac{f(a) f^T(a)}{a}
\end{equation*}
can be expressed in the form
\be
\label{C-1}
C_a^{-1}=M_a^{-1} - \frac{M_a^{-1} f(a) f^T(a) M_a^{-1}     }{ a+ f^T(a) M_a^{-1} f(a)}    \, .
\ee
We now discuss the cases $f(0) \neq 0 $ and $f(0) = 0 $ separately.

\begin{theorem}
\label{thm4}
Consider the continuous time linear regression model \eqref{mod1cont} on the interval $[0,b]$
with a continuously differentiable vector $f$ of regression functions.
If each component of $f$ is of bounded variation, $1 \notin {\rm span}\{f_1, \ldots, f_m\} $ and $f(0) \neq 0 \in \mathbb{R}^m$,  then the estimator
\begin{equation}\label{asblue0}
\hat{\theta}_{\rm BLUE}   = \underline{C} \,  \int_0^b \dot{f}(t) \,dY_t + \frac{M_0^{-1} f(0)    }{ f^T(0) M_0^{-1} f(0)}   Y_0  \, ,
\end{equation}
is the best linear unbiased estimator,
where
\begin{equation*}
\underline{C}= \lim_{a \to 0} C_a^{-1} =M_0^{-1} - \frac{M_0^{-1} f(0) f^T(0) M_0^{-1}     }{ f^T(0) M_0^{-1} f(0)}  =  \mathrm{Var} (\hat{\theta}_{\rm BLUE} )   \, .
\end{equation*}
\end{theorem}

{\bf Proof.}
For any $a>0$  the BLUE  $\hat{\theta}_{{\rm BLUE},a}$
 in the continuous time model \eqref{mod1cont} on the interval $[a,b]$
 is given by \begin{equation}\label{asblueA}
\hat{\theta}_{{\rm BLUE},a}   = C_a^{-1} \Big( \int_a^b \dot{f}(t) \,dY_t + \frac{ f(a)}{a} Y_a \Big) .
\end{equation}
As
$a\to 0$,
\bea
\lim_{a\to 0} C_a^{-1}  \int_a^b \dot{f}(t) \,dY_t = \underline{C} \int_0^b \dot{f}(t) \,dY_t
\eea
and
\bea
\lim_{a \to 0}
C_a^{-1} \frac{ f(a)}{a} &=   &
\lim_{a \to 0}  \Big( M_a^{-1} \frac{ f(a)}{a} - \frac{M_a^{-1} f(a) f^T(a) M_a^{-1}  f(a)   }{ a(a+ f^T(a) M_a^{-1} f(a)) } \Big) \\
& = &
\lim_{a \to 0}  \frac{ M_a^{-1}f(a)}{a+ f^T(a) M_a^{-1} f(a))} =\frac{M_0^{-1} f(0)    }{ f^T(0) M_0^{-1} f(0)}
\eea
Hence the left-hand side of \eqref{asblue0} is the limit of the estimators
$\hat{\theta}_{{\rm BLUE},a}$ as $a \to 0$. The  covariance matrix of this estimator is obtained
by Ito's formula and the fact that $\varepsilon_0=0$ , i.e.
\bea
\mathrm{Var} (\hat{\theta}_{\rm BLUE})
& =& \underline{C} \left[ \int_0^b \dot{f}(t) \dot{f}^T(t)\,d t\,\right] \underline{C} = \underline{C} M_0 \underline{C} =
 I - \frac{M_0^{-1} f(0) f^T(0)      }{ f^T(0) M_0^{-1} f(0)}  \underline{C}=\underline{C}\, .
\eea
In order to prove that the derived estimator \eqref{asblue0}  is in fact BLUE we use  Theorem 2.3 in \cite{N1985a}, which
states that an unbiased estimator  of the form $\hat{\theta}= \int_a^b Y_t dG(t)$ with covariance matrix
 $C= {\rm Var}(\hat{\theta}) $  is BLUE in model \eqref{eq:model}  if the identity
\be
\label{eq:W-H}
\int_a^b K(s,t)dG(s)=C f(t)
\ee
holds for all $t\in [a,b]$.
Here $G$ is a vector measure on the interval $[a,b]$. In the present case $a=0$
and  $K(s,t)=\min (s,t)$, and in order to prove that the estimator
\eqref{asblue0} is indeed BLUE we use  the representation
\begin{equation*}
\int_0^b \dot{f}(t) \,dY_t = \dot{f}(b) Y_b - \dot{f}(0) Y_0 - \int_0^b  Y_t d \dot{f} (t) ,
\end{equation*}
for the stochastic integral $\int_0^b \dot{f}(t) \,dY_t$. This defines the vector measure $dG$ in an obvious manner,
 i.e. it has mass  $\underline{C} \dot{f}(b)$ at the point  $b$, the density $-\underline{C} \ddot{f}(t)$
for $t \in [0,b]$ and some mass at the point $0$. The validity of \eqref{eq:W-H} for $\hat{\theta}_{\rm BLUE}$ and $\underline C$ now
follows from
\bea
- \int_0^b \min (s,t) d  \dot{f} (s)   &=&  - \int_0^t s d  \dot{f} (s) - t \int_t^b  d  \dot{f} (s) \\ &=&- [t\dot{f}(t)-f(t)+f(0)]-t[ \dot{f}(b)-\dot{f}(t)]=-f(0)+ f(t) - t \dot{f}(b),
\eea
by noting that $\underline{C} {f}(0)=0$ and that the weight at $b$ cancels out.
 \hfill $\Box$

 \bigskip

If $f(0) = 0 \in \mathbb{R}^m$, the observation at $t=0$ necessarily gives $Y_0=0$
and provides no further information about the parameter $\theta$. We obtain the following result.

\begin{theorem}
\label{thm5}
Consider the continuous time linear regression model \eqref{mod1cont} on the interval $[0,b]$
with a continuously differentiable vector $f$ of regression functions.
If each component of $f$ is of bounded variation, $1 \notin {\rm span}\{f_1, \ldots, f_m\} $ and $f(0) = 0 \in \mathbb{R}^m$,
then
\begin{equation}\label{asblue1}
\hat{\theta}_{\rm BLUE}   = M_0^{-1} \,  \int_0^b \dot{f}(t) \,dY_t  \, ,
\end{equation}
and
\begin{equation*}
\mathrm{Var} (\hat{\theta}_{\rm BLUE} )
= M_0^{-1}
\end{equation*}
\end{theorem}
{\bf Proof.}
Since for any $p=1,\ldots,m$ the function  $f_p(t)$ is continuously differentiable on $[0,b]$, the limit $ \lim_{t \to 0} f_p(t)/t  $ is necessarily finite, possibly 0.
Using this and  the fact that $f(0)=0$,  the representation \eqref{C-1} gives
$
\lim_{a \to 0} C_a^{-1}= M_0^{-1},
$
and  the limit of $\hat{\theta}_{{\rm BLUE},a} $ defined in \eqref{asblueA}
is obviously \eqref{asblue1}. The
 covariance matrix of this  estimator is again obtained by an application of Ito's formula  and its
 optimality follows by similar arguments as given in the proof of Theorem \ref{thm4}.
\hfill $\Box$

\subsection{Models with an  intercept, that is  $1 \in {\rm span}\{f_1, \ldots, f_m\} $}

W.l.o.g. we may assume $f_1(t)=1$ for all $t \in [0,b]$ and rewrite
 the original regression model \eqref{mod1cont}  as
\bea
Y_t=\theta_1+ \tilde{\theta}^T \tilde{f}(t) + \varepsilon_t,~~\quad t\in[0,b],
\eea
where
$\tilde{\theta}=(\theta_2, \ldots, \theta_m)^T$ and $\tilde{f}(t)=( {f}_2(t), \ldots,  {f}_m(t))^T)$.
Note that  the observation at $t=0$ is error-free and gives
$
Y_0=\theta_1+ \tilde{\theta}^T \tilde{f}(0).
$
By subtracting  we obtain
\be
\label{mod4}
Y_t-Y_0= \tilde{\theta}^T ( \tilde{f}(t)- \tilde{f}(0) ) + \varepsilon_t .
\ee
Note that
$1 \notin {\rm span}\{\tilde{f}_2(t)-\tilde f_2(0), \ldots, \tilde{f}_m(t)-\tilde f_m(0)\} $
and $\tilde{f}(t)-\tilde{f}(0) $ is obviously 0 at $t=0$.
For computing the BLUE  for $\tilde \theta$ and its covariance matrix  in model \eqref{mod4} we can apply Theorem~\ref{thm5}
and obtain
\begin{eqnarray}
\tilde{\theta}_{\rm BLUE}  &=& \tilde{M}_0^{-1} \,  \int_0^b \dot{\tilde{f}}(t) \,d(Y_t)  \, ,
\\
\mathrm{Var} (\tilde{\theta}_{\rm BLUE} )
&=&  \tilde{M}_0^{-1} = \left[ \int_0^b \dot{\tilde{f}}(t) \dot{\tilde{f}}^T(t) dt \right]^{-1} \, .
\end{eqnarray}
Finally, the BLUE for $\theta_1$ is given by
$
\hat{\theta}_1=Y_0-\tilde{\theta}_{\rm BLUE}^T \tilde{f}(0).
$
Noting that $Y_0$ is a constant, we obtain
$
{\rm cov }(\hat{\theta}_1,\hat{\theta}_p) = -\tilde{f}^T(0) M_0^{-1} e_p\;\;\;(p=2, \ldots, m)\, ,
$
where $e_p$ is the $p$-th coordinate vector. The variance  of $\hat{\theta}_1$ is given by ${\rm Var }(\hat{\theta}_1)=  \tilde{f}^T(0) M_0^{-1} \tilde{f}(0)  $.

\bigskip

{\bf Acknowledgements.}
This work has been supported in part by the Collaborative
Research Center ``Statistical modeling of nonlinear dynamic processes'' (SFB 823, Teilprojekt C2) of the German Research Foundation (DFG).
The research of H. Dette reported in this publication was also partially supported by the National Institute of
General Medical Sciences of the National Institutes of Health under Award Number R01GM107639.
The content is solely the responsibility of the authors and does not necessarily
 represent the official views of the National
Institutes of Health.
We would also like to thank Kirsten Schorning for her constructive comments on an earlier version of this manuscript and
 Martina Stein who typed parts of this paper with considerable technical expertise.
Parts of this paper have been written while the authors were visiting the Isaac Newton Institute, Cambridge, and we would like to thank the institute for its hospitality.

\setlength{\bibsep}{1pt}
\bibliography{opt_signed_designsa}

\begin{appendix}
 \section{Appendix: More technical details}

\subsection{An auxiliary result} \label{aux}

\begin{lemma}
\label{l:2 cases}
Let $f(t) =(f_1(t), \ldots, f_m(t))^T $ be a vector of continuously differentiable linearly independent functions on
the interval  $[a,b]$ with $0 \leq a<b$ and define
$
M = \int^b_a \dot{f} (s) \dot{f}^T(s)ds .
$
\begin{enumerate}
  \item
The matrix $M$ is non-singular  if and only if $1 \notin {\rm span}\{f_1, \ldots, f_m\} $.
  \item  If $1 \in {\rm span}\{f_1, \ldots, f_m\} $
then ${\rm rank} (M)=m-1$.
\end{enumerate}

\end{lemma}

{\bf Proof.}  \\
(1)  Obviously the non-singularity of $M$ implies that $1 \notin {\rm span}\{f_1, \ldots, f_m\} $. To prove the converse we consider
 the equation
\be
\label{eq:lin_dep}
 a_1\dot{f_1}(t)+ \ldots a_m \dot{f_m}(t)=0, \;\; \forall t \in [a,b]
\ee
for scalars $ a_1,\ldots, a_m$. This equation
is satisfied if and only if for some $a_0$ we have
\be
\label{eq:lin_dep1}
  a_0+ a_1{f_1}(t)+ \ldots a_m {f_m}(t)=0, \;\; \forall t \in [a,b]\, .
\ee
By the assumption,
the functions ${f_1}, \ldots, {f_m}$ are linearly independent on  the interval $[a,b]$ and   $1 \notin {\rm span}\{f_1, \ldots, f_m\} $, which implies
that  the $m+1$ functions $1,{f_1}, \ldots, {f_m}$
are also linearly independent on $[a,b]$. Consequently
the equation \eqref{eq:lin_dep1} has only the trivial solution $a_0=a_1=\ldots=a_m=0$.
which yields that the equation \eqref{eq:lin_dep} has only trivial solution $a_1=\ldots=a_m=0$.
Therefore  the
 functions $\dot{f_1}(t), \ldots, \dot{f_m}(t)$ are linearly independent on the interval
 $[a,b]$ and the non-singularity of $M$ follows from basic results on Gramian matrices [see  \cite{akhgla1981}, p. 18].
 \smallskip

 (2) To prove the second part assume  now that  $1 \in {\rm span}\{f_1, \ldots, f_m\} $. Since  $f_1, \ldots, f_m$ are linearly
 independent we may assume w.l.o.g. that $f_1(t)= {\rm const}$ for all $t \in [a,b]$. In this case, $\dot{f_1}=0$ and
 $1 \notin {\rm span}\{f_2, \ldots, f_m\} $ and part (1) shows that
the $(m-1)\times (m-1)$  submatrix of the matrix  $(\int_a^b  {f_k}(s) {f_l}(s) ds)_{k,l=2,\ldots ,m}$
has full rank, which implies that  ${\rm rank} (M)=m-1$.
\hfill $\Box$

\subsection{Proof of Lemma \ref{lemma0}} If  $1 \notin {\rm span}\{f_1, \ldots, f_m\} $ if follows  from Lemma~\ref{l:2 cases} in Section \ref{aux}
that the matrix
$M$ is non-singular and hence positive definite, which implies $C>0$.
If  $1 \in {\rm span}\{f_1, \ldots, f_m\} $ we may assume w.l.o.g. that $f_1(t) \equiv 1$. As the
functions $f_2,\ldots, f_m$ are linearly independent and  $1  \notin {\rm span}\{f_2,\ldots f_m\}$ it follows that
$$
M = \int_a^b \dot{f}(t) \dot{f}^T(t) dt =\left(
        \begin{array}{cc}
          0 & 0 \\
          0 & \tilde{M}
        \end{array}
      \right)
$$
where (by  Lemma~\ref{l:2 cases}) the matrix $\tilde M= ( \int_a^b \dot{f}_k(t) \dot{f}_l^T(t) dt  )_{k,l=2}^m$ has rank $m-1$. Define
 $f(t)= (1 , \tilde{f}(t)^T), $ where $\tilde{f}^T (t)= (f_2, \ldots, f_m)$ and
 assume that the matrix
$ C$ is singular. Then there exists a vector $z= (z_1, \tilde{z}^T) \in \mathbb{R}^{m} \setminus \{0\}$
with  $\tilde{z} \in \mathbb{R}^{m-1}$
such that
\bea
z^T C z = z^T M z + \frac{z^T f(a) f^T(a)z}{a}=  \tilde{z}^T \tilde{M} \tilde{z} + (z^T f(a))^2/a\,  =0 .
\eea
As both terms in the sum are nonnegative  we have $\tilde{z}^T \tilde{M} \tilde{z}=0$ and $z^T f(a)=0$. Since $\tilde{M}$ is a positive definite matrix
we obtain $\tilde{z}=0 \in \mathbb{R}^{m-1}$. The equation $z^T f(a)=0$ then becomes $z_1 f_1(0)=0$ implying $z_1=0$ and
 hence $z= 0 \in \mathbb{R}^m$. This yields a contradiction to the assumption that the matrix  $ C$ is singular and proves Lemma \ref{lemma0}.
\hfill $\Box$

\subsection{Proof of Lemma \ref{criterion-multi-dimension}} \label{a3}
  Define the random variables
$$
X_i = \int_{t_{i-1}}^{t_i} [\dot{f}(s) - \mu_i] \,dY_s, \qquad
  i=2, \ldots, n.
  $$  From the definition of  $\hat{\theta}_{\rm BLUE}  $ and $ \hat{\theta}_n$ in \eqref{asblue} and \eqref{estimatormul}, respectively, we have
\begin{align*}
\mathbb{E}_\theta[(\hat{\theta}_{\rm BLUE}  - \hat{\theta}_n)(\hat{\theta}_{\rm BLUE}  - \hat{\theta}_n)^T]
&= C^{-1} \mathbb{E}_\theta \Big[ \sum_{i=2}^n X_i \sum_{j=2}^n X_j ^T \Big]  C^{-1}.
\end{align*}
Observing the fact that the random variables  $X_2,\ldots ,X_n$ are independent
we obtain
\begin{equation*}
\mathbb{E}_\theta \Big[ \sum_{i=2}^n X_i \sum_{i=2}^n X_i ^T \Big] = \sum_{i=2}^n \mathbb{E}_\theta \big [( X_i -\mathbb{E}_\theta [X_i] )(  X_i-\mathbb{E}_\theta [X_i] )^T]  +
\sum_{i=2}^n \mathbb{E}_\theta [X_i]  \sum_{j=2}^n \mathbb{E}_\theta [X_j ^T]  .
\end{equation*}
Ito's isometry yields
\begin{equation*}
\mathbb{E}_\theta[X_i] = \int_{t_{i-1}}^{t_i} [\dot{f}(s) - \mu_i] \dot{f}^T(s) \theta ds,  \qquad i=2,\ldots,n,
\end{equation*}
 and
\begin{align*}
\mathbb{E}_\theta [(X_i -\mathbb{E}_\theta [X_i] )(X_i-\mathbb{E}_\theta [X_i] )^T] &=
 \mathbb{E}_\theta \Big[   \int_{t_{i-1}}^{t_i} [\dot{f}(s) - \mu_i] \, d \varepsilon_s
  \int_{t_{i-1}}^{t_i} [\dot{f}(s) - \mu_i] ^T\, d \varepsilon_s  \Big] \\
&= \int_{t_{i-1}}^{t_i} [\dot{f}(s) - \mu_i] [\dot{f}  (s) - \mu_i ]^T \,ds .
\end{align*}
Therefore,
\begin{align*}
\mathbb{E}_\theta  \Big[ \sum_{i=2}^n X_i \sum_{i=2}^n X_i ^T \Big]
&= \sum_{i=2}^n \int_{t_{i-1}}^{t_i} [\dot{f}(s) - \mu_i] [\dot{f} (s) - \mu_i ]^T \,ds\\
&+ \sum_{i=2}^n \int_{t_{i-1}}^{t_i} [\dot{f}(s) - \mu_i] \dot{f}^T(s) \theta \,ds \sum_{j=2}^n \int_{t_{j-1}}^{t_j} \theta^T \dot{f}(s)[\dot{f}(s) - \mu_j ]^T \,ds,
\end{align*}
which proves the assertion. \hfill $\Box$

\subsection{Proof of Theorem \ref{equivalence}.} \label{a4}
Standard calculations show that
$$
\mathbb{E}_\theta [\hat \theta_n] = C^{-1} \Big[ \sum^n_{i=2} \mu_i (f(t_i)-f(t_{i-1}))^T + \frac {f(a)f^T(a)}{a} \Big] \theta.
$$
 Observing the definition of the matrix $C$ in \eqref{cmatrix1} it follows that the estimator $\hat{\theta}_n$ defined in \eqref{estimator} is unbiased if and only if the identity \eqref{unbiasmul} is satisfied.
In order to prove the second part of Theorem \ref{equivalence} we
 use the decomposition
  \begin{align} \label{decomp}
\mathbb{E}_\theta[(\tilde{\theta}_n - \theta)(\tilde{\theta}_n - \theta)^T] &= E_1 + E_2 + E_2^T + E_3,
\end{align}
where the terms $E_1,$ $E_2$  and  $E_3$ are defined by
\begin{align*}
E_1 &=
\mathbb{E}_\theta [ (\tilde{\theta}_n - \hat{\theta}_{\rm BLUE} )(\tilde{\theta}_n - \hat{\theta}_{\rm BLUE} )^T]  ,  \\
E_2 &=   \mathbb{E}_\theta [(\tilde{\theta}_n - \hat{\theta}_{\rm BLUE} )(\hat{\theta}_{\rm BLUE}  - \theta)^T ] , \\
E_3 &=  \mathbb{E}_\theta [ (\hat{\theta}_{\rm BLUE}  - \theta)(\hat{\theta}_{\rm BLUE}  - \theta)^T].
\end{align*}
By Theorem \ref{thm1} we have
\begin{equation*}
E_3 =  C^{-1} = \left[ \int_a^b \dot{f}(s) \dot{f}^T(s) \,ds + \frac{f(a) f^T(a)}{a} \right]^{-1}.
\end{equation*}
Using   the definition of $\tilde \theta_n $ and $\hat \theta_{\rm BLUE} $  in  \eqref{asblue},
yields
\begin{align*}
C (\tilde{\theta}_n - \hat{\theta}_{\rm BLUE} )
&= C \int_{a}^{b} g(s) \,dY_s - \int_a^b \dot{f}(s) \,dY_s -\frac{f(a)}{a} Y_a\\
&= C \int_{a}^{b} g(s) \dot{f}^T(s) \theta \,ds + C \int_{a}^{b} g(s) \,d \varepsilon_s - \int_a^b \dot{f}(s) \dot{f}^T(s) \theta \,ds - \int_a^b \dot{f}(s) \,d \varepsilon_s \\
&- \frac{f(a) f^T(a)}{a} \theta - \frac{f(a)}{a} \varepsilon_a \\
&=  \int_{a}^{b} [C g(s) - \dot{f}(s)] \, d \varepsilon_s -\frac{f(a)}{a} \varepsilon_a ,
\end{align*}
where the last identity follows from the fact that $\tilde{\theta}_n$ is unbiased, that is,
\begin{equation}\label{313new}
\int_a^b g(s) \dot{f}^T(s) ds = I.
\end{equation}
On the other hand
\begin{align*}
C (\hat{\theta}_{\rm BLUE}  - \theta) &= \int_a^b \dot{f}(s) \,dY_s + \frac{f(a)}{a} Y_a - \int_a^b \dot{f}(s) \dot{f}^T(s) \,ds \theta - \frac{f(a) f^T(a)}{a} \theta \\
&= \int_a^b \dot{f}(s) \, d\varepsilon_s  + \frac{f(a)}{a} \varepsilon_a .
\end{align*}
Therefore we obtain for the term $E_2$ the representation
\begin{align*}
E_2  &=C^{-1} \Big\{ \mathbb{E}_\theta\Big[ \Big( \int_{a}^{b} [C g(s) - \dot{f}(s)] \, d \varepsilon_s -\frac{f(a)}{a} \varepsilon_a \Big)
\Big( \int_a^b \dot{f}(s) \,d\varepsilon_s + \frac{f(a)}{a} \varepsilon_a  \Big)^T \Big] \Big\} C^{-1} \\
&= C^{-1} \Big\{ \mathbb{E}_\theta \Big[ \int_{a}^{b} [C g(s) - \dot{f}(s)] \, d \varepsilon_s  \int_a^b \dot{f}^T(s) \,d \varepsilon_s  \Big] - \mathbb{E}_\theta \Big[ \frac{f(a)}{a} \varepsilon_a \varepsilon^T_a \frac{f^T(a)}{a} \Big] \Big\} C^{-1} \\
&= C^{-1} \Big [\int_{a}^{b} [C g(s) - \dot{f}(s)] \dot{f}^T(s) \, ds - \frac{f(a) f^T(a)}{a}  \Big] C^{-1} \\
&= C^{-1} \Big [ C- \int_a^b \dot{f}(s) \dot{f}^T(s) \, ds - \frac{f(a) f^T(a)}{a} \Big] C^{-1} = 0 ,
\end{align*}
where the last identity is again a consequence of \eqref{313new}.
Hence it follows from \eqref{decomp}
\begin{equation*}
\mathbb{E}_\theta[(\tilde{\theta}_n - \theta)(\tilde{\theta}_n - \theta)^T] = \mathbb{E}_\theta[(\tilde{\theta}_n - \hat{\theta}_{\rm BLUE} )(\tilde{\theta}_n - \hat{\theta}_{\rm BLUE} )^T] + C^{-1},
\end{equation*}
which proves the assertion of Theorem \ref{equivalence}.
\hfill $\Box$

\end{appendix}
\end{document}